# The 802.11 MAC Protocol Leads to Inefficient Equilibria

Godfrey Tan and John Guttag
MIT Computer Science and Artificial Intelligence Laboratory
Cambridge, MA 02139
{godfreyt, guttag}@csail.mit.edu

*Abstract*— Wireless local area networks (WLANs) based on the family of 802.11 technologies are becoming ubiquitous. These technologies support multiple data transmission rates. Transmitting at a lower data rate (by using a more resilient modulation scheme) increases the frame transmission time but reduces the bit error rate. In non-cooperative environments such as public hot-spots or WLANs operated by different enterprises that are physically close to each other, individual nodes attempt to maximize their achieved throughput by adjusting the data rate or frame size used, irrespective of the impact of this on overall system performance.

In this paper, we show both analytically using a game theoretic model and through simulation that the existing 802.11 distributed MAC protocol, DCF (for Distributed Coordination Function), as well as its enhanced version, which is being standardized at part of 802.11e, can lead non-cooperative nodes to undesirable Nash equilibriums, in which the wireless channel is inefficiently used. We show that by establishing independence between the allocation of the shared channel resource and the transmission strategies used by individual nodes, an ideal MAC protocol can lead rational nodes to arrive at equilibriums in which all competing nodes achieve higher throughputs than with DCF.

## I. INTRODUCTION

This paper deals with issues related to the way in which 802.11 WLANs resolve contention for the channel in non-cooperative environments, such as public hot-spots or private enterprises that are physically close to each other (e.g. neighboring office suites in a commercial building or neighboring residences). In such environments, multiple nodes may compete for channel access in a *rational* but *non-cooperative* manner. That is each competing node will maximize its utility regardless of what other nodes achieve.

The performance of indoor wireless systems is affected by many factors, including the location of transmitters and receivers and the complex characteristics of indoor RF channels. During congested periods, contention among nodes plays a particularly important role. The bandwidth of the 2.4GHz band used by 802.11 is wide enough for 3 orthogonal 802.11b or 802.11g channels, i.e., a maximum of 3 nodes can simultaneously transmit data with little interference. Contention among nodes using the same channel is resolved using DCF, a variant of CSMA (Carrier Sense Multiple Access) protocol As the deployment of WLANs continues to grow rapidly, we believe that neighboring wireless networks that are managed by different administrative authorities will often need to share a common channel due to scarcity of available spectrum.

In indoor environments, the channel loss rates of nodes vary widely: even receiving nodes that are equi-distant from a common sender experience differing channel conditions [6]. When the average signal strength at the receiver is lower than the threshold required for successful frame reception, the sender can unilaterally elect to use an alternative coding scheme that exploits the trade-off between data rate and BER [4]. Transmitting at a lower data rate by using a more resilient modulation scheme leads to higher frame transmission time but reduces the frame loss rate.

Figure 1 shows the achieved TCP throughput of a sender as a function of the distance between it and a receiver in a simulated environment. The channel model used in this simulation is described in detail in Section IV. For each pair of data rates, there exists a *cross-over distance* at which using a lower data rate yields higher throughput because the reduction in frame loss rate at the lower data rate is high enough to compensate for the slower transmission speed. For instance, at distances greater than 100 m, transmitting at 5.5 Mbps yields higher throughputs than transmitting at 11 Mbps.

In a rational world, the cross-over distance defines the optimal transmission rate for senders. But the world of 802.11 and DCF is not quite rational in this respect.

Each competing 802.11 node can use any 802.11-compliant strategy to maximize its achieved throughput. An 802.11 node can determine, for each frame transmission, the frame size and the data transmission rate. If each competing node uses the most efficient transmission strategy, i.e., the strategy that yields the highest achievable throughput when the node alone occupies the channel, the resulting aggregate throughput will be optimal.

However, in the presence of competition under DCF, a rational node may not use its most efficient transmission rate. The root cause of this behavior is the mechanism used by DCF to dictate how the medium is shared. DCF is designed to give an approximately equal probability of channel access (measured in number of *transmission opportunities* (TXOPs)) to each competing node with similar loss characteristics. That is to say, over any period lasting hundreds of milliseconds, each node will be able to transmit an equal number of frames, *irrespective of the amount of channel time required to transmit the frame*.

A rational node will attempt to maximize its throughput by maximizing the product of its share of channel occupancy time and its achievable throughput per unit of occupancy. Under DCF, the share of channel occupancy time a node obtains depends on the data rates used by it and its competitors. By intentionally transmitting at a lower data rate, a node may achieve a higher channel time share than it would by transmitting at a higher and more efficient data rate. This effect combined with the reduction in the node's frame loss rate may lead to higher achieved throughput for that node. However, this is done at the cost of overall efficiency. The aggregate throughput lost by other nodes will exceed the throughput gained by the node using a non-optimal transmission rate.

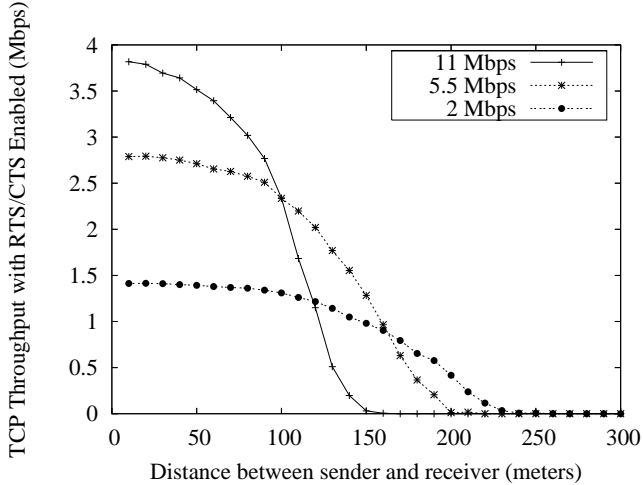

Fig. 1. TCP throughputs achieved at various data rates in a simulated environment.

If a node has the channel all to itself, it will never use a data rate lower than its optimal data rate, since doing so would result in reduced throughput. However, in the presence of another competing node, that node may transmit at a lower data rate, since by doing so it can use the channel longer and experience a lower loss rate. As a result, the overall efficiency of the network significantly suffers.

In this paper, we:

- Develop a game theoretic model for rational wireless nodes competing for the shared channel resource,
- Show analytically using our model and through simulation that under certain conditions both DCF and its enhanced cousin EDCF (for Enhanced Distributed Co-ordinating Function), which is being drafted as part of 802.11e, can lead competing rational nodes to undesirable equilibriums, in which the shared wireless channel is inefficiently used, and
- Show that by guaranteeing the allocation of long-term shares of channel time to competing nodes with respect to a desired fairness constraint, the MAC protocol can force rational nodes to efficiently use the shared medium, thereby improving the combined achieved throughputs of all competing nodes.

## II. RELATED WORK

Previous studies [3], [12] have discussed some undesirable effects that DCF has on overall network performance when multiple competing nodes use different data rates. For instance, when competing nodes transmit at different data rates, the aggregate throughput will be dominated by the lowest transmission rate. Tan *et al.* [12] explain in detail the differences between *time-based fairness*, in which each node is given an equal amount of channel time, and *throughput-based fairness*, in which each node achieves equal throughputs.

The work discussed above argues for changing the definition of fairness to achieve increased aggregate throughput. Our work is independent of the choice of fairness criterion. We analyze the conditions under which DCF allows rational non-cooperative nodes to use inefficient strategies. The observations and results shown in later sections do not depend on a particular notion of fairness. Our arguments are based solely on efficiencies and not on fairness. To our knowledge, we are the first to show that both DCF and EDCF lead rational nodes, each of which chooses the data rate that yields the maximum achieved throughput, to undesirable equilibria.

## III. ANALYSIS

For simplicity, in this section we conduct our analysis on UDP flows. TCP complicates the analysis since we need to take into account frame loss rates in both directions, one for TCP data packets and the other for TCP ack packets. However, our simulation results in Section IV show that the end results of our theorems also hold for TCP flows.

### A. Network Model

Let $i$ be a mobile sender and let $I$ be the set of senders contending for channel access. Let $T$ be the duration, in seconds, during which competing nodes continuously wish to send UDP data. For node $i$, we define the following terms that characterize its communication process during $T$:

- $d_i$: the average data rate in Mbps,
- $s_i$: the average payload size (per-frame) in bits,
- $t_i$: the channel occupancy time in seconds, and
- $\alpha(s_i, d_i, I)$: the overall frame success rate as a fraction.

The success rate $\alpha(s_i, d_i, I)$ is a function of the data rate and frame size used and the level of contention and channel conditions experienced. $\alpha(s_i, d_i, I)$ is simply 1 minus the average frame loss rate. Generally, losses due to channel errors decrease with the reduction in frame size or data rate. Losses due to collisions decrease with the decreased level of contention. For UDP, the steady-state UDP throughput taking into account MAC-layer losses can be characterized by $\alpha(s_i, d_i, I)$. We also assume that there are no losses due to buffer overflows. In Section IV, we report on a simulation using a channel model that reflects typical channel conditions in indoor mobile environments. Its results are consistent with the analysis of this section.

We define the *channel occupancy time $t_i$*, of node $i$ as the number of seconds node $i$ used for transmitting frames during $T$. The channel occupancy time necessary to transfer each data frame includes i) the transmission time of the data frame, ii) the transmission time of a synchronous MAC-layer *ack*, which is transmitted by the receiver 10 microseconds after successfully receiving the data frame, iii) the propagation delays, and iv) the time, such as inter-frame idle periods, necessary for a node to be idle before accessing the channel. Note that each retransmitted frame adds to the channel occupancy time used. As we discuss later, under DCF, $t_i$ is a function of the data rate and frame size used.

In a typical WLAN, even if a node has more demand than the capacity, it will not be able to use 100% of the channel time for frame transmission. This is largely due to the inefficiency of any CSMA-style distributed MAC protocol that requires competing nodes to remain idle for random intervals to avoid collisions. Let $t_{idle}$ be the idle time when no nodes access the channel during $T$, then

$$T = \sum_{i \in I} t_i + t_{idle} \qquad (1)$$

We define $\gamma(d_i, s_i)$, the theoretically achievable throughput as follows:

$$\gamma(s_i, d_i) = \frac{s_i}{to_i + \frac{s_i + bo_i}{d_i}} \qquad (2)$$

where $bo_i$ is the bit overhead and $to_i$ the time overhead. The bit overhead is the number of bits in the frame that are not application payload bits, such as MAC and transport protocol headers. The time overhead is the combined time necessary to transmit a physical layer preamble, the synchronous acknowledgment and the interframe space time between the data and ack frames. For given data rate $d_i$ and frame size $s_i$, $\gamma(d_i, s_i)$ is the upper bound for the achieved throughput of node $i$ and can be obtained both theoretically and experimentally. In general, node $i$'s achieved throughput will be much less than $\gamma(d_i, s_i)$ because of idle periods and frame losses. Observe that since $bo_i$ and $to_i$ are constants, $\gamma(s_i, d_i)$ increases with increased $s_i$ and $d_i$.

We define the practically achievable throughput $R^{prac}$ as:

$$R^{prac}(d_i, s_i) = \gamma(d_i, s_i) * \alpha_i(d_i, s_i, I) \qquad (3)$$

$R^{prac}$ is the upper bound for node $i$'s achieved throughput under given channel conditions. It only depends on the efficiency of the MAC protocol and the channel conditions experienced by the node. If a node is rational and has the channel all to itself, it will only select a pair of data rate and frame size that maximizes $R^{prac}$.

We use $d^{max}$, $d^{min}$, and $s^{max}$ to denote the maximum data rate, the minimum data rate and the maximum frame size respectively. For concreteness, we based our analysis on the simple yet general scenario of two nodes $i$ and $j$ sending UPD packets. They compete for channel access and employ their best local strategies to maximize achieved throughputs.

All nodes are assumed to be within radio range of each other. We note that the analysis also holds true without this assumption, for the following reasons. When some nodes are hidden from other nodes, the overall performance of a DCF-style MAC protocol can significantly degrade because of increased collisions. Traditionally, this *hidden* terminal problem is addressed through a virtual carrier sense mechanism such as the RTS(request-to-transmit)/CTS(clear-to-transmit) mechanism [1]. When the RTS/CTS protocol is used, each transmission opportunity will be preceded by an exchange of a pair of packets, an RTS packet and a CTS packet. The overhead of transmitting RTS and CTS packets can be reflected in $to_i$ and thus the analysis of DCF is not impacted by the use of the RTS/CTS mechanism for environments where hidden nodes exist. Our simulation results using the RTS/CTS protoco support our assertion as we show in Section IV

### B. Game Model

In this section, we model two rational, non-cooperative nodes, $i$ and $j$, each sending UDP data to a receiver as two players playing a finitely repeated non-cooperative game. In each stage, stagegame $Gm$ is played as follows. The first node transmits a burst of $b_i \leq n$ frames successively. Following that, the second node transmits a burst of up-to $m$ frames successively. Under our assumption that nodes always have frames to transmit, each node will attempt to transmit the maximum numbers of frames allowed. However, the actual number of frames transmitted ($b_i$) may be less than the maximum allowed depending on the backoff technique used by the MAC protocol.

A stagegame may last no more than $\tau$ seconds. At the beginning of each stagegame, with probability $p$ node $i$ communicates first and with probability $1-p$ node $j$ communicates first. For the rest of the paper, we assume that $p = 0.5$ and consider a $K$-*repeated* game $Gm(K)$ in which the stagegame $Gm$ is played $K$ times and $K$ is even. The values of $n$, $m$ and $\tau$ are dictated by the underlying MAC protocol.

Again, the utility of each player is its achieved UDP throughput over $\tau * K$ seconds. At each stagegame, the available actions of each player are to set its data rate and to set its frame size. The goal of each competing player is to employ the strategy $g^* = (d^*, s^*)$ that maximizes its achieved throughput given the other player's best transmission strategy.

### C. Nash and Subgame Perfect Equilibriums

In each stagegame $Gm$, nodes are in a Nash Equilibrium (NE) if each node does not have any incentive to deviate from its current strategy of using a specific combination of data rate and frame size. Note that there could be more than one NE in each stagegame.

An outcome of a $K$-*repeated* game $Gm(K)$ is the achieved throughputs of the two nodes given their strategies over all $K$ stagegames. To simplify our analysis, we assume that the overall channel conditions remain relatively unchanged. In other words, the success probability of frame transmission observed by node $i$ over each interval of $\tau$ seconds in each stagegame is similar. While this assumption is typically valid for relatively static environments where channel errors occur randomly, it is not valid for mobile environments in which a moving sender or receiver can lead to correlated frame losses on a short timescale [7], [10]. That is channel conditions of some stagegames may drastically differ from that of other stagegames in mobile environments. However, our simulation results show that our analysis still holds in mobile environments.

A *subgame* beginning at stage $k+1$ of $Gm(K)$ is the repeated game in which stagegame $Gm$ is played $K-k$ times and is denoted $Gm(K-k)$. An outcome of a $K$-*repeated* game ($Gm(K)$) is considered *subgame perfect* if in each subgame, only NE strategies are played. Note that in general there could be many *subgame perfect equilibriums (SPEs)* for a $K$-*repeated* game since there could be more than one NEs. However, if stagegame $Gm$ has a unique NE, then the finitely repeated game $Gm(K)$ also has a unique SPE, in which the unique NE of $Gm$ is played at every stagegame [2]. In other words, if the competing nodes are rational, each node will use the unique transmission strategy (the strategy used in the unique NE) in each stagegame of $Gm(K)$.

Ideally, each node should use a strategy that yields the maximum practically achievable throughput, leading to the maximum aggregate achieved throughput with respect to the particular channel time allocation. Therefore, an outcome in which each node employs a strategy yielding the maximum achievable throughput is considered *desirable*. A NE is considered *desirable* if its outcome is desirable and otherwise is considered *undesirable*. Similarly, a SPE of a $K$-*repeated* game is *desirable* if a desirable NE is played at each subgame and otherwise *undesirable*.

In non-cooperative environments, a rational player $i$ may use a strategy that yields non-optimal practically achievable throughput but achieves a higher time share ($Frac_i$), thereby, achieving higher throughput. As a result, one or more undesirable NEs may exist in the stagegame. Nonetheless, when there exists at least one desirable NE in the stagegame, a desirable SPE (for the $K$-*repeated* game) may still be reached since rational nodes can use threats of retaliation to force a desirable SPE [2]. However, when the stagegame has a unique NE and that NE is undesirable, the resulting unique SPE of the $K$-*repeated* game is also undesirable.

The rest of this section shows that DCF, in many occasions, and EDCF, in some occasions, lead rational nodes to arrive at undesirable unique NEs (and thus undesirable unique SPEs). Naturally, one might ask whether it is possible to design the MAC protocol so that it can always lead to desirable SPEs in non-cooperative environments. We show in Sections III-F and V that this is indeed possible.

### D. Achieved Throughput

The achieved steady-state UDP throughput of a wireless node $i$ employing strategy $g_i = (d_i, s_i)$ when competing against node $j$ employing $g_j = (d_j, s_j)$ during $T(g_i, g_j) = t_i(g_i, g_j) + t_j(g_i, g_j) + t_{idle}$ is:

$$R_i(g_i, g_j) = R^{prac}(g_i) * f_i(g_i, g_j), \quad (4)$$

where

$$f_i(g_i, g_j) = \frac{t_i(g_i, g_j)}{t_i(g_i, g_j) + t_j(g_i, g_j) + t_{idle}} \quad (5)$$

The achieved throughput is a product of $R^{prac}$ and $f_i(g_i, g_j)$, the fraction of channel occupancy time obtained by node $i$. When multiple nodes share a common channel, the actual achieved throughput of the node will depend on the fraction of the channel occupancy time it gets. The underlying MAC protocol greatly influences the values of $t_i$ and $t_j$.

For two strategies, $g_1 = (d_1, s_1)$ and $g_2 = (d_2, s_2)$, where $d_1 > d_2$ and $s_1 = s_2$, the following properties of the underlying physical and MAC layer protocols are assumed:

*Statement 1:* $\gamma(g_1) > \gamma(g_2)$
Clearly, the theoretically achievable throughput at a higher data rate is larger than that at a lower data rate as evident in Equation 2.

*Statement 2:* $\alpha_i(g_1) \leq \alpha_i(g_2)$
The loss rate at a lower data rate is at least as good as that at a higher data rate. This statement is true for various physical layer coding schemes used by a family of 802.11 technologies [4]. Since all our analyses involve only two nodes, we will use $\alpha_i(g_1)$ instead of $\alpha_i(d_1, s_1, I)$.

*Statement 3:* $t_{idle}$ (in each round) remains constant for any competition involving two nodes using any strategies.
The (average) per-frame overhead for winning a transmission opportunity depends mainly on the number of nodes contending for the channel access (not on the transmission strategies used).

### E. Analysis of DCF

DCF gives an equal long-term channel access probability to each contender with similar channel conditions [5], [13]. However, when two nodes experiencing different loss rates compete, the long-term channel access probability of the node with the higher loss rate will be lower. This is caused by the backoff algorithm that forces a node to backoff longer whenever it experience a failed transmission. Our results for DCF hold regardless of the existence of such an artifact. For simplicity, we ignore this artifact. Thus, we assume that under DCF, competing nodes sending data frames over the same time interval will be able to transmit approximately equal numbers of frames. Note that DCF only allows a single frame to be transmitted during each transmission opportunity. Therefore, when nodes use DCF, we can specify the game as follows: $b_i = b_j = n = m = 1$.

In the rest of this section, we prove theorems and claims using concrete examples and intuitions. More rigorous and formal arguments for these claims and theorems can be found in the Appendix.

*Lemma 1:* Under DCF, the amount of time each node achieves during each stagegame is the amount of time required to transmit its data frame using its transmission strategy. I.e., $t_i(g_i, g_j) = \frac{s_i}{\gamma(g_i)}$ and $t_j(g_i, g_j) = \frac{s_j}{\gamma(g_j)}$.

*Proof:* Since each node only gets to transmit one frame each in a stagegame ($b_i = b_j = 1$), the claim is self-evident. ∎

*Theorem 1:* Under certain channel conditions, there exists undesirable unique SPEs under DCF.

*Proof:* We construct a concrete example illustrating the existence of an undesirable unique SPE. Assume that there are two data rates $d_1$ and $d_2$ and that $d_1 > d_2$ and that $t_{idle} = 0$. Also, assume that each node uses maximum-sized frames and thus there are only two viable strategies that each node can choose: $g_1 = (d_1, s^{max})$ and $g_2 = (d_2, s^{max})$. Furthermore, assume that the channel conditions for node $i$ and $j$ are as follows:

| Strategy | $\gamma(g_x)$ | $\alpha_i(g_x)$ | $\alpha_j(g_x)$ |
|---|---|---|---|
| $g_1$ | 3.2 | 0.6 | 1 |
| $g_2$ | 1.6 | 0.95 | 1 |

According to the table, node $j$ suffers no losses at any data rate. However, node $i$ experiences a higher success rate when transmitting at the lower data rate $d_2$.

Based on Lemma 1, Equations 4, 3 and 5, we can compute all the possible outcomes of each stagegame as follows:

| $g_i \setminus g_j$ | $g_1$ | $g_2$ |
|---|---|---|
| $g_1$ | (0.96, 1.6) | (0.63, 1.07) |
| $g_2$ | **(1.02, 1.06)** | (0.76, 0.8) |

Each pair represents the achieved throughputs of the two nodes each using the given strategies. For instance, the top left pair $(0.96, 1.6)$ denotes the achieved throughputs of node $i$ and node $j$ respectively, given that node $i$ uses strategy $g_1$ and node $j$ uses strategy $g_1$. From this table, it is clear that there exists a unique NE in which node $i$ plays strategy $g_2$ and node $j$ plays strategy $g_1$.

And this unique NE is undesirable. It is easy to see that if node $i$ is the only one transmitting, the most efficient strategy is clearly $g_1$, i.e., $\gamma(g_1) * \alpha_i(g_1) > \gamma(g_2) * \alpha_i(g_2)$. Similarly, $g_1$ is the most efficient strategy for node $j$. Unfortunately, at the unique NE, node $i$ uses a less efficient strategy $g_2$. As a result, the aggregate throughput at equilibrium $(1.02 + 1.06 = 2.08)$ is less than the aggregate throughput that could have achieved $(0.96 + 1.6 = 2.76)$ if both nodes use their most efficient strategies.

Since the stagegame has the unique NE, the finitely repeated game also has the unique SPE [2]. Also observe that there are many sets of channel conditions that can lead to undesirable unique NEs. For instance, if $\alpha_i(g_2) = 0.92$, an undesirable unique SPE will ensue. In Section IV, we show that these situations are common using a realistic channel model for mobile environments. ∎

The fundamental problem with DCF is that it treats transmission opportunities as the common resource to be shared and allocated. As we demonstrated earlier, this treatment is not suitable in the presence of multiple data transmission rates, frame size and different channel conditions among competing nodes. Providing fixed long-term channel access probabilities while allowing variable channel time per transmission opportunity (as DCF does) leads rational nodes to use inefficient transmission strategies since they can increase their channel time shares by doing so.

*F. Analysis of EDCF*

In an attempt to provide QoS support for 802.11-based WLANs, an IEEE working group is drafting the 802.11e standard that specifies a distributed channel access protocol, EDCF, an enhancement to DCF. Unlike DCF, EDCF allows a node that wins the contention to transmit for a bounded interval of time $t^{max}$, irrespective of the frame size and data rate used. It appears that the main reason for limiting the duration of each TXOP is the predictability of the maximum frame transmission time, which is necessary to meet QoS guarantees. This limit also significantly affects the nature of competition.

EDCF, unlike DCF, allows bursts of frames to be transmitted. The maximum burst length depends on the data rate used. For instance, for $t^{max} = 7.35$ ms, at least five 1500-byte frames can be successively transmitted at 11 Mbps. However, at 5.5 Mbps, only about three 1500-byte frames can be transmitted. Like DCF, EDCF gives an equal long-term channel access probability (i.e., equal number of TXOPs) to competing nodes that have the same priority. However, the actual number of frames transmitted by a node in a transmission opportunity (on average) depends on the backoff scheme.

Distributed MAC protocols like DCF and EDCF employ a backoff scheme to resolve contention. Under DCF, after each frame transmission, a node picks a random number of 20-$\mu$s time slots between 0 and the contention window size ($cw$) and remains idle during that backoff period. This allows another contender with a smaller backoff period to access the channel. Inevitably, frames sometimes collide and the number of collisions increases rapidly with the number of contenders. DCF uses an exponential backoff technique in which the contention window size is doubled for each failed frame transmission. If the previous frame transmission is successful, $cw$ is set to a pre-determined minimum value, $cw_{min}$.

Under EDCF, a node can transmit multiple frames per transmission opportunity and any of those frames can be lost. The time at which a node backs off can affect the amount of channel time it gets. There are two major ways in which this can be done.

First, a node can stop transmitting subsequent frames as soon as it detects a failed transmission within the burst. We call this technique BFL (for Backoff upon First Loss). Since the wireless channel is lossy, the average of number of frames transmitted per transmission opportunity typically will be lower than the maximum allowed. Subsequently, the average channel time used per stagegame will be less than the maximum allowed, i.e., $t_i < t^{max}$.

Second, a node can transmit the maximum number of frames allowed regardless of failures, and backs off only after the last frame transmission. We call this technique BEB (for Backoff at End of Burst). Under BEB, the average number of frames transmitted per transmission opportunity will be equivalent to the maximum allowed, i.e., $t_i = t^{max}$.

There are advantages and disadvantages to each technique. When there is only a single node transmitting, it is better to employ BEB since it increases the achieved throughput by reducing the total amount of backoff time ($t_{idle}$). However, when multiple nodes are competing for channel access and losses are bursty, BFL is more desirable than BEB. In indoor mobile environments, channel conditions are time-correlated on short time scales because of multipath and mobility [10], and thus, whenever a frame transmission fails due to channel errors, it is likely that successive frame transmissions will also fail. Thus, under BFL, a node will avoid likely failed transmissions by backing off as soon as it experiences a frame loss. Meanwhile, a competing node with better channel conditions can transmit, improving the overall efficiency. It has been observed that the channel qualities of different transmission paths are often independent and thus losses on a single path are often bursty in mobile environments [7], [11]. As we explain shortly, EDCF with BFL leads rational nodes to use inefficient equilibrium strategies but EDCF with BEB does not.

*Lemma 2:* Under EDCF, $t_i(g_i, g_j) = \frac{b_i * s_i}{\gamma(g_i)}$ and $t_j(g_i, g_j) = \frac{b_j * s_j}{\gamma(g_j)}$.

*Proof:* The channel time allocated to node $i$ under DCF during each stagegame is simply the amount of channel occupancy time needed to transmit the average number of frames ($b_i$) transmitted in each stagegame. ∎

*Theorem 2:* Under EDCF with BFL, there exists undesirable unique SPEs.

*Proof:* We construct an example illustrating the existence of an undesirable unique SPE. Assume that there are two data rates $d_1$ and $d_2$ and that $d_1 > d_2$ and that $t_{idle} = 0$. Also, assume that each node uses maximum-sized frames and thus there are only two viable strategies that each node can choose: $g_1 = (d_1, s^{max})$ and $g_2 = (d_2, s^{max})$. Assume also that $t^{max} = 0.015$ seconds and $s^{max} = 1500$ bytes. Thus, each node can transmit a maximum of 4 frames using $g_1$ and a maximum of 2 frames using $g_2$. Furthermore, assume that the channel conditions of node $i$ and $j$ are as follows:

| Strategy | $\gamma(g_x)$ | $\alpha_i(g_x)$ | $b_i(g_x)$ | $\alpha_j(g_x)$ | $b_j(g_x)$ |
|---|---|---|---|---|---|
| $g_1$ | 3.2 | 0.6 | 2.18 | 1 | 4 |
| $g_2$ | 1.6 | 0.95 | 1.95 | 1 | 2 |

We note that under EDCF with BFL, $b_i$ is really the expected number of transmissions in each stagegame and depends on the transmission strategy used and the channel conditions experienced. $b_i(g_x)$ can also be computed from the overall frame loss rate, $\alpha_i$, as follows provided that the loss process is random (as we assume). Let $p_i(k)$ be the probability of node $i$ transmitting a $k$ frames in each stagegame.

$p_i(k) = \alpha_i^{(k-1)} * (1 - \alpha_i)$ and
$b_i = \sum_{k=1}^{n-1}(p_i(k) * k) + (1 - \sum_{k=1}^{n-1} p_i(k)) * n_i$

Recall that $n$ is the maximum number of frames that node $i$ can transmit in each stagegame. For instance, the expected number of transmissions at each stagegame for node $i$ using $g_1$ is $b_i(g_1) = 0.4 * 1 + 0.6 * 0.4 * 2 + 0.6 * .0.6 * 0.4 * 3 + 0.216 * 4 = 2.18$.

Based on Lemma 2 and Equations 3, 4 and 5, we can compute all possible outcomes of each stagegame as follows:

| $g_i \setminus g_j$ | $g_1$ | $g_2$ |
|---|---|---|
| $g_1$ | (0.68, 2.07) | (0.68, 1.04) |
| $g_2$ | **(0.75, 1.62)** | (0.75, 0.81) |

Again, each pair represents the achieved throughputs of the two nodes, each using a given strategy. From this table, it is clear that there exists a unique NE in which node $i$ plays strategy $g_2$ and node $j$ plays strategy $g_1$. Unfortunately, this unique NE is undesirable. It is easy to see that if node $i$ is the only one transmitting, the most efficient strategy is clearly $g_1$, i.e., $\gamma(g_1) * \alpha_i(g_1) > \gamma(g_2) * \alpha_i(g_2)$. Unfortunately, at the unique NE, node $i$ uses a less efficient strategy $g_2$. As a result, the aggregate throughput at equilibrium $(0.75 + 1.62 = 2.37)$ is less than the aggregate throughput that could have been achieved $(0.68 + 2.07 = 2.75)$ if both nodes used their most efficient strategies. ∎

*Theorem 3:* Let $g_i^*$ and $g_j^*$ be the strategies of nodes $i$ and $j$ at a NE. Under EDCF with BEB, $g_i^*$ and $g_j^*$ are desirable strategies. I.e., any NE arrived under EDCF with BEB is desirable. Since any NE is desirable, any SPE will also be desirable.

Intuitively, the theorem states that if the system allocates the same amount of channel time regardless of the strategy used, each node at equilibrium will always use the strategy that yields the maximum practically achievable throughput.

*Proof:* We prove by contradiction. Suppose that there exists a strategy, $g_i' \neq g_i^*$, such that $R^{prac}(g_i') > R^{prac}(g_i^*)$. Since $t_i^* = t^{max}$, $t_i' \leq t_i^*$. Thus, according to Equation 4, $R_i(g_i', g_j^*) > R_i(g_i^*, g_j^*)$. But this contradicts the fact that $g_i^*$ is the optimal equilibrium strategy for node $i$, given that node $j$ uses $g_j^*$. A similar argument can be made for $g_j^*$ being a desirable strategy.

However, as explained in Section III-F, EDCF with BEB can lead to higher overall frame loss rates than EDCF with BFL. In other words, the aggregate throughputs achieved in a SPE can be improved if the MAC protocol provides flexibility. In Section V, we show how such an adaptive MAC protocol can gain the advantages of both BFL and BEB, leading to improved aggregate throughputs. ∎

## IV. SIMULATION

We conduct simulation runs in $ns$ [8], relaxing some of the simplifying assumptions made in our analytical model.

### A. Environments

We use a Rayleigh fast-fading model [9], [10] to capture the short time-scale fading phenomenon that arises because of objects moving along the transmission path between a transmitter and a receiver, which may also be moving. The received power thresholds for various data rates are based on the Orinoco 802.11b Gold Card data sheet.

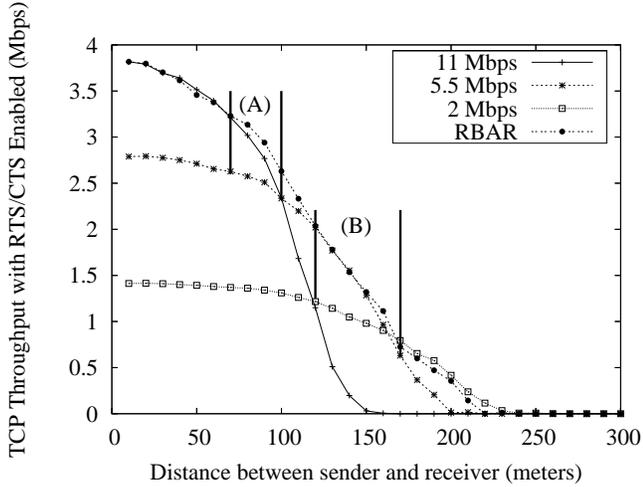

Fig. 2. TCP throughputs achieved when using various fixed data rates and RBAR, an auto-rate protocol. Regions (A) and (B) are where rational nodes under DCF may use inefficient strategies when competing against nodes with lower loss rates (smaller transmission distances).

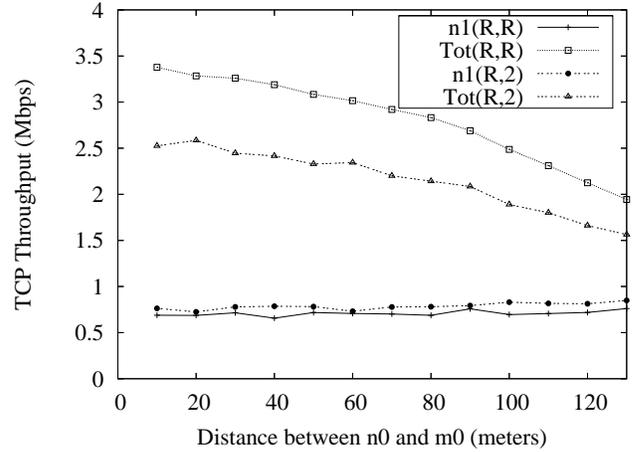

Fig. 4. TCP throughputs achieved by $n1$ and the aggregate achieved throughputs under two pairs of strategies as a function of the distance between $n0$ and $m0$. $(R, 2)$ denotes that $n0$ uses RBAR and $n1$ transmits at a fixed data rate of 2 Mbps. $Tot(R, 2)$ plots the aggregate throughputs. However, the most efficient strategy for $n1$ is to transmit at 5.5 Mbps, which is what RBAR running at $n1$ would do. Thus, $(R, R)$ denotes the most efficient strategy pair which may not be used at equilibriums.

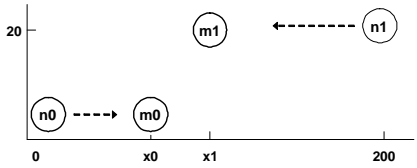

Fig. 3. $n0$ and $n1$ transmit to $m0$ and $m1$ respectively.

Unlike in the previous section, we use TCP instead of UDP, to demonstrate that our results apply to TCP. Also in our analytical model, we also assumed that the channel conditions in each subgame are constant, leading to each node transmitting at the most appropriate transmission rate for the entire duration, given that all other nodes choose their best transmission rates. In practice, channel conditions vary and wireless card vendors employ proprietary auto-rate adaptation schemes that adjust the data transmission rate (on a frame-by-frame basis) based on estimated channel conditions. Our simulation takes into account auto-rate protocols. As we show in this section that our analytical results agree with the simulation results of more realistic scenarios.

For concreteness in our examples, we use the Receiver-based Auto-rate protocol (RBAR) [4] although our results do not depend on a particular auto-rate protocol. An RBAR receiver informs a sender of channel conditions before the sender transmits a data frame. In particular, the sender sends an RTS (request to transmit) frame and the receiver reports the received signal strength of the RTS frame in a replying CTS (clear to transmit) frame. The RTS/CTS scheme is typically used to reduce collisions as a result of frame transmissions by hidden nodes. Compared to data frames, the RTS and CTS frames are very small and are transmitted at 2 Mbps making them robust against channel errors. Based on the signal strength information, the sender then chooses the highest transmission rate at which successful frame transmission is highly likely, under the assumption that the channel conditions will remain unchanged for the transmission period. Figure 2 shows that in most cases RBAR performs well as it adapts the transmission rate based on observed channel conditions.

However, a rational node may not choose its transmission strategy solely based on its channel conditions. In practice, a rational node will periodically evaluate its achieved throughput, channel conditions, observed channel time usage and average frame loss rate to determine the best strategy for transmitting data frames. Such a scheme can be practically implemented at the MAC layer, but is beyond the scope of this paper. As evident in our analyses in the previous sections and this section, the best transmission strategy that maximizes the achieved throughput of an individual node is not necessarily the most efficient one.

*B. Results*

We ran experiments using the setup shown in Figure 3. There are two TCP flows, one from $n0$ to $m0$ and the other from $n1$ to $m1$. Note that $m0$ and $m1$ also send TCP acknowledgment packets to $n0$ and $n1$ respectively. The positions of $n0$, $m1$, and $n1$ were fixed whereas that of $m0$ was varied. $m1$ was 130 m away from $n1$ (i.e., $x1 = 70$ m), and the distance between $n0$ and $m0$ was varied from 10 to 130 m. All nodes are within radio transmission range of each other.

We also ran a set of experiments using UDP flows. The results were similar in nature and since TCP is most widely used, we only include the results for TCP experiments.

When both nodes used RBAR, $n1$ achieved lower throughput than $n0$ when its distance from $m1$ was farther than that between $n0$ and $m0$. Notice that the most efficient data rate for $n1$ would be 5.5 Mbps if $n1$ had the channel all to itself (see Figure 2). In fact, this was what RBAR did most of the time. However, in the presence of a competing flow, $n1$ could achieve higher throughput by transmitting at 2 Mbps. This behavior is evident in Figure 4 which shows the achieved throughputs of $n1$ and the aggregate throughputs as a function of the distance between $n0$ and $m0$. For example, when $m0$ is 10 m away from $n0$, $n1$ can achieve an 11% increase in throughput by always transmitting at a lower data rate instead of using RBAR. However, as a result of $n1$ using this inefficient strategy, the achieved throughput of $n0$ (not shown in the figure) and the aggregate throughput would decrease by 53% and 34% respectively.

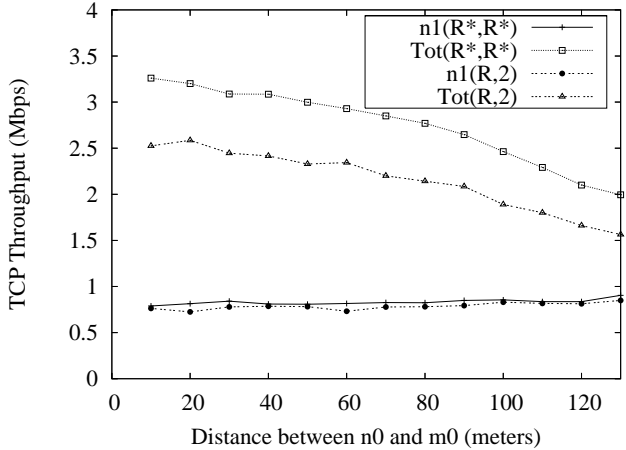

Fig. 5. TCP throughputs achieved by $n1$ and the aggregate throughputs as the distance between $n0$ and $m0$ varies. As explained in Figure 4, under DCF, $n1$ gains the highest throughputs by transmitting at a less efficient rate of 2 Mbps. $n1(R,2)$ and $Tot(R,2)$ plot the achieved throughputs of $n1$ and the aggregate throughputs under DCF. Under the hypothetical MAC protocol $DCF^*$, using $(R^*,R^*)$ leads to the highest achieved throughputs for both nodes. $Tot(R*,R*)$ plots the aggregate throughputs under $DCF^*$, which are superior to those achieved under DCF.

In Figure 4, $n1$ only gains an 11% increase in throughput by transmitting at a less efficient rate of 2 Mbps instead of transmitting at 5.5 Mbps. However, the figure only shows an example scenario illuminating the impact of arriving at inefficient equilibriums under DCF. There are certainly cases where $n1$ could gain much higher throughputs by transmitting at inefficient data rates at the expense of reducing aggregate throughputs.

We ran numerous experiments to determine the regions in which rational nodes could benefit by transmitting at an inefficient data rate. In Figure 2, a node in region $A$ or $B$ can achieve higher throughput by choosing a data rate lower than the most efficient data rate, whenever it competes against node that experiences a lower loss rate. The wide ranges of regions $A$ and $B$ highlight the importance of incorporating mechanisms to reduce inefficiencies as a result of competition among rational nodes in non-cooperative environments.

The simulation results (not shown here) for EDCF with FLB are similar to those described here although the regions where rational nodes may use inefficient strategies under EDCF are smaller than those under DCF.

## V. ACHIEVING EFFICIENT EQUILIBRIUMS

As stated in Theorem 3, any NE arrived under EDCF with BLB is desirable. That is to say the equilibrium strategy of each competing node is the same as the one it would have used if it were the only node using the channel. However, as explained in Section III-F, BLB is inflexible in that it does not allow potential optimization techniques to reduce frame loss rates. When losses are bursty, as they often are in mobile environments, a node can reduce its loss rate by not transmitting subsequent frames (that are highly likely to lead to failures) as soon as a frame loss is experienced. Meanwhile, another competing node which has better channel conditions can transmit. Such an optimization technique has been proposed for centralized, cooperative WLANs [7] and is shown to reduce frame loss rates by as much as 25%.

In non-cooperative environments, a rational node has little incentive to give up its channel time if it does not receive some compensation (in the future). We suggest, therefore, that the MAC protocol guarantees each competing node that it will achieve its assigned channel time share over a designated time period no matter how much channel time is used per transmission opportunity. We propose that the ideal MAC protocol should:

- Determine the desired long-term allocation of the shared medium based on channel time, instead of transmission opportunities[1],
- Limit the amount of channel time used per transmission opportunity (for the purpose of avoiding starvation), i.e., $t_i \leq t^{max}$, and
- Dynamically allocate the probability of transmission opportunities as a function of the observed channel time share so that the observed long-term global allocation of channel time is not affected by the transmission strategies used by nodes. I.e., $\sum_{x=1}^{K} t_i^x$ should remain unchanged irrespective of the transmission strategy used.

### A. How to Achieve Efficient Equilibriums

We use the following example to illustrate i) the potential realizable gains when nodes use the most desirable transmission strategies, and ii) how we might go about providing the long-term time-share guarantees at the MAC layer. We use the same example provided in Section IV-B. Under DCF, $n1$ transmits at 2 Mbps even though 5.5 Mbps is the data rate with the highest practically achievable throughput if $n1$ had the channel all to itself.

To entice $n1$ to employ the most efficient transmission strategy, a MAC protocol, $DCF^*$, could give $n1$ a higher probability of channel access than it would normally get under DCF by lowering its contention window size, $cw_{min}$. Recall that under DCF, $cw_{min} = 31$ for both $n0$ and $n1$. Under $DCF^*$, we set $cw_{min} = 26$ for $n1$ and $cw_{min} = 36$ for $n0$. Figure 5 shows that under $DCF^*$ the achieved throughput of $n1$ using RBAR, $n1(R^*, R^*)$, is always higher than what $n1$ achieved by always transmitting at 2 Mbps under DCF, $n1(R, 2)$. Moreover, the aggregate throughput under $DCF^*$ is more than 30% higher than that under DCF. Note that under $DCF^*$, $n1$ would get the same amount of channel time regardless of its transmission strategy. Therefore $n1$, being rational, will not intentionally lower its data rate under $DCF^*$.

---
[1]The MAC layer should have a default policy of the proportions of channel time that each competing node should achieve. E.g., equal time shares for competing nodes is a reasonable notion.

We also note that $DCF^*$ will encourage nodes to employ optimization techniques to avoid burst losses. This in turn increases the overall efficiency of the network in terms of throughputs and one-way delay.

We believe that the MAC protocol for future 802.11 standards should provide the properties outlined earlier so that rational 802.11-compliant devices use efficient strategies. There are two major challenges in developing such a scheme. First, the MAC protocol running at each node must observe its share of channel time. The information needed to do this is already available under DCF. The MAC protocol at each node not only knows its channel time used but is also aware of the channel time used by each neighboring node via existing physical and virtual carrier sensing mechanisms. Second, the MAC protocol at each node must periodically determine its contention window size as a function of its channel time share. This must be done in a way that ensures the convergence of the observed global channel time allocation over a per-determined period to the desired allocation. We plan to implement such a scheme. We believe that the increased protocol complexity will be more than offset by realizable significant performance gains in non-cooperative environments, but have yet to demonstrate this experimentally.

## VI. CONCLUSIONS

The rapid growth of independently managed wireless local area networks leads to increasing competition for the wireless channel among wireless devices. The MAC layer protocols in use today do not prevent rational nodes from behaving in ways that degrade aggregate network performance. Through extensive game theoretical analyses as well as through simulation, we showed that under certain conditions, both DCF and its (future) enhanced cousin EDCF force rational non-cooperative nodes to use inefficient strategies at unique NEs, leading to significantly reduced network throughput.

We showed through simulation that long-term channel time share guarantees (rather than transmission opportunity guarantees) can be used to ensure that rational competing nodes use the channel time allocated to them in the most efficient manner. A MAC protocol can achieve this goal by dynamically adjusting in a distributed manner the contention window size of each node as a function of its observed channel time share.

Our approach takes into the practicalities of the marketplace. For each 802.11 wireless interface card, there are two functional components: *standard-compliant and customizable*. The *standard-compliant* component includes implementations (usually in firmware) of MAC and physical layers that are compliant with IEEE 802.11 specification. Thus, parameters such as $CW_{min}$ should be set according to the specification. In practice, each 802.11 product undergoes a certification process administered by the Wi-Fi Alliance, a nonprofit international association formed in 1999 to certify interoperability of WLAN products based on IEEE 802.11 specification [14]. Presumably, the certification process will verify whether a product is compliant with the specification. Assuming that 802.11 wireless interface manufacturers want a wide-acceptance of their products by being standard-compliant and certified products, there is little incentive for them to improve performance of their products in a way that violates the specification. For instance, in theory, a node may opt to transmit frames without backing off, i.e., set $cw_{min} = 0$. But a rational manufacturer may not do that with the fear its products not being certified and hurting its reputation.

On the other hand, each manufacturer or even user can customize a lot of MAC layer related parameters that are left unspecified by the standard. We consider data rate and frame size as part of the *customizable component* since 802.11 specification does not limit how such parameters are used. In fact, in practice, each card manufacturer often has its own proprietary auto-rate protocol to choose an appropriate data rate for each frame transmission, as we mentioned before. Furthermore, users can also adjust those parameters by modifying publicly available software drivers that act as the interface between the private firmware implementation of the MAC protocol and the networking stack of the operating system. As we have demonstrated throughout the paper, enhancements to the 802.11 MAC protocol is necessary to prevent rational nodes from arriving at inefficient equilibriums by modifying customizable parameters such as data rate.

## ACKNOWLEDGEMENTS

The authors would like to thank Rahul Sami for giving useful comments for our camera-ready version.

## APPENDIX

*Analysis of DCF*

**Lemma 4 and Lemma 5 serve as a formal proof of Theorem 1 which we informally prove in Section III-E by giving an example.**

*Claim 1:* Under DCF, for any three strategies $g_1 = (d_1, s_1)$, $g_2 = (d_2, s_2)$, and $g_3 = (d_3, s_3)$, where $d_1 > d_2$ and $s_1 = s_2 = s_3 = s$, $f_i(g_1, g_3) < f_i(g_2, g_3)$.

*Proof:* According to Lemma 1, $t_i(g_1, g_3) < t_i(g_2, g_3)$ and $t_j(g_1, g_3) = t_j(g_2, g_3)$. Based on Equation 5, we can see that $f_i(g_1, g_3) < f_i(g_2, g_3)$.

*Claim 2:* Under DCF, for any three strategies $g_1 = (d_1, s_1)$, $g_2 = (d_2, s_2)$, and $g_3 = (d_3, s_3)$, where $d_1 > d_2$ and $s_1 = s_2 = s_3 = s$, $\gamma_{(g_1)} * f_i(g_1, g_3) > \gamma_{(g_2)} * f_i(g_2, g_3)$.

*Proof:* According to Lemma 1 and Equation 5, $\gamma_{(g_1)} * f_i(g_1, g_3) = \frac{s}{t_i(g_1,g_3)+t_j(g_1,g_3)+t_{idle}}$ and $\gamma_{(g_2)} * f_i(g_2, g_3) = \frac{s}{t_i(g_2,g_3)+t_j(g_2,g_3)+t_{idle}}$. Since $t_j(g_1, g_3) = t_j(g_2, g_3)$ and $t_i(g_1, g_3) < t_i(g_2, g_3)$ (as evident by Lemma 1), $\gamma_{(g_1)} * f_i(g_1, g_3) > \gamma_{(g_2)} * f_i(g_2, g_3)$.

*Claim 3:* Under DCF, if node $i$ experiences no losses when transmitting at the highest data rate $d^{max}$ using the maximum frame size $s^{max}$, the strategy $g_i^* = (d^{max}, s^{max})$ is the *dominant* strategy of node $i$, i.e., $\forall g_i \neq g_i^*$ and $\forall g_j$, $R_i(g_i^*, g_j) > R_i(g_i, g_j)$.

*Proof:* We have $\forall d_i \neq d^*, d^* > d_i$. Thus, according to Lemma 2, $\forall g_i \neq g_i^*$ and $\forall g_j$, $\gamma(g_i^*) * f_i(g_i^*, g_j) > \gamma(g_i) * f_i(g_i, g_j)$. Since $\alpha_i(g_i^*) = 1$, $\forall g_i \neq g_i^*$ and $\forall g_j$, $\gamma(g_i^*) * f_i(g_i^*, g_j) * \alpha(g_i^*) > \gamma(g_i) * f_i(g_i, g_j) * \alpha(g_i)$ (i.e., $R_i(g_i^*, g_j) > R_i(g_i, g_j)$).

*Claim 4:* DCF can lead to a unique subgame perfect equilibrium in which a unique NE is played at each stage game.

*Proof:* We show that by construction. Assume that both nodes use maximum-sized frames. Also assume that node $j$ has the dominant strategy $g_j^*$, i.e., $\forall g_i$ and $\forall g_j \neq g_j^*$, $R_j(g_i, g_j^*) > R_j(g_i, g_j)$. $(g_i^*, g_j^*)$ forms a unique NE if $\forall g_i \neq g_i^*$, $\frac{\alpha_i(g_i^*)}{\alpha_i(g_i)} > \frac{T(g_i^*, g_j^*)}{T(g_i, g_j^*)}$. According to Lemma 1 and Equations 3, 4 and 5, it is easy to see that satisfying this condition leads to $R_i(g_i^*, g_j^*) > R_i(g_i, g_j^*)$.

Note that the condition $\frac{\alpha_i(g_i^*)}{\alpha_i(g_i)} > \frac{T(g_i^*, g_j^*)}{T(g_i, g_j^*)}$ is easily satisfied if $g_i^*$ involves using the highest data rate $d^{max}$ and $\alpha_i(g_i^*) = 1$. However, in general, that is not the only case. For example, even if $d_i^* < d^{max}$ and $\alpha_i(g_i^*) < 1$, the above condition can still hold. Without loss of generality, assume that $d^{max} > d_i^* > d^{min}$. For $g_i = (d^{max}, s^{max})$, it's possible that $\frac{\alpha_i(g_i^*)}{\alpha_i(g_i)} > \frac{T(g_i^*, g_j^*)}{T(g_i, g_j^*)}$: the right hand side is greater than 1 (since $t_i(g_i, g_j^*) < t_i(g_i^*, g_j^*)$) but $\alpha_i(g_i^*)$ can be greater than $\alpha_i(g_i)$ (see Statement 1).

For $g_i = (d^{min}, s^{max})$, it's possible that $\frac{\alpha_i(g_i^*)}{\alpha_i(g_i)} > \frac{T(g_i^*, g_j^*)}{T(g_i, g_j^*)}$. The right hand size is less than 1 (since $t_i(g_i, g_j^*) > t_i(g_i^*, g_j^*)$). As long as $\alpha_i(g_i)$ is not much higher than $\alpha_i(g_i^*)$, $g_i^*$ can be the dominant strategy. In conclusion, the dominant strategy $g_i^*$ can constitute any data rate (not just the highest data rate).

*Claim 5:* Let there be two possible pairs of strategies $(g_i', g_j^*)$ and $(g_i^*, g_j^*)$ where $g_i' \neq g_i^*$. Furthermore, let $d_i' > d_i^*$ and $s_i'=s_i^*=s_j^*=s$. If $g_i^*$ and $g_j^*$ are the unique NE strategies under DCF, the NE may be undesirable (and as a result, the subgame perfect equilibrium is also undesirable).

Informally, this claim states that if node $i$ and node $j$ use the same frame size and node $i$ is not transmitting at the fastest data rate at equilibrium (i.e., there exists $d_i' > d_i^*$), the strategies at the unique NE may be inefficient.

*Proof:* We prove by showing that for certain combinations of $\alpha_i(g_i^*)$ and $\alpha_i(g_i')$, it is possible for node $i$ to employ $g_i^*$ as an equilibrium strategy even though $g_i'$ yields higher practically achievable throughput, i.e., $R^{prac}(g_i') > R^{prac}(g_i^*)$.

Since $(g_i^*, g_j^*)$ are the unique Nash Equilibrium strategies,

$R_i(g_i^*, g_j^*) > R_i(g_i', g_j^*)$

According to Equation 4 and the given assumptions,

$R^{prac}(g_i^*) * f_i(g_i^*, g_j^*) > R^{prac}(g_i') * f_i(g_i', g_j^*)$

According to Equation 3 and Lemma 1,

$\gamma(g_i^*) * \alpha_i(g_i^*) * \frac{\frac{s}{\gamma(g_i^*)}}{T(g_i^*, g_j^*)} > \gamma(g_i') * \alpha_i(g_i') * \frac{\frac{s}{\gamma(g_i')}}{T(g_i', g_j^*)}$

$\frac{T(g_i', g_j^*)}{T(g_i^*, g_j^*)} > \frac{\alpha_i(g_i')}{\alpha_i(g_i^*)}$

Also, according to Equation 1,

$\frac{T(g_i', g_j^*)}{T(g_i^*, g_j^*)} = \frac{\frac{s}{\gamma(g_i')} + \frac{s}{\gamma(g_j^*)} + t_{idle}}{\frac{s}{\gamma(g_i^*)} + \frac{s}{\gamma(g_j^*)} + t_{idle}}$

$= \frac{\gamma(g_i^*)}{\gamma(g_i')} * \frac{s*\gamma(g_i') + s*\gamma(g_j^*) + t_{idle}*\gamma(g_i')*\gamma(g_j^*)}{s*\gamma(g_i^*) + s*\gamma(g_j^*) + t_{idle}*\gamma(g_i^*)*\gamma(g_j^*)}$

Therefore,

$\frac{\gamma(g_i^*)}{\gamma(g_i')} * \frac{s*\gamma(g_i') + s*\gamma(g_j^*) + t_{idle}*\gamma(g_i')*\gamma(g_j^*)}{s*\gamma(g_i^*) + s*\gamma(g_j^*) + t_{idle}*\gamma(g_i^*)*\gamma(g_j^*)} > \frac{\alpha_i(g_i')}{\alpha_i(g_i^*)}$

Since $\gamma(g_i^*) < \gamma(g_i')$ (because $d_i' > d_i^*$), the second left term is greater than 1.

So, it is possible that

$\frac{\gamma(g_i^*)}{\gamma(g_i')} < \frac{\alpha_i(g_i')}{\alpha_i(g_i^*)}$

$R^{prac}(g_i^*) < R^{prac}(g_i')$ (see Equation 3)

Intuitively, node $i$ will use a less efficient data rate $d^*$ as the equilibrium strategy instead of a more efficient strategy $d'$ so long as the proportional increases in the success rate of frame transmission and in the channel times allocated is higher than the proportional reduction in achievable throughput.

*Analysis of EDCF*

**Lemmas 6 and 7 serve as a formal proof of Theorem 2, which we informally prove in Section III-F.**

*Claim 6:* EDCF (with FLB or BLB) can lead to a unique subgame perfect equilibrium in which a unique NE is played at each stage game.

*Proof:* We prove it by construction. Assume that both nodes use maximum-sized frames. Also assume that node $j$ has the dominant strategy $g_j^*$, i.e., $\forall g_i$ and $\forall g_j \neq g_j^*$, $R_j(g_i, g_j^*) > R_j(g_i, g_j)$. $(g_i^*, g_j^*)$ forms a unique NE if $\forall g_i \neq g_i^*$, $\frac{\alpha_i(g_i^*)}{\alpha_i(g_i)} > \frac{b_i * T(g_i^*, g_j^*)}{b_i^* * T(g_i, g_j^*)}$. According to Lemma 2 and Equations 3, 4 and 5, it is easy to see that this condition leads to $R_i(g_i^*, g_j^*) > R_i(g_i, g_j^*)$.

One example scenario where the necessary condition holds is when $g_i^* = (d^{max}, s^{max})$, $g_j^* = (d^{max}, s^{max})$, $\alpha(g_i^*) = \alpha(g_j^*) = 1$. Note that this claim can also hold true in many scenarios where $d_i^* \neq d^{max}$ and $\alpha(g_i^*) < 1$ for reasons similar to those given in Lemma 4.

*Claim 7:* Let there be two possible pairs of strategies $(g_i', g_j^*)$ and $(g_i^*, g_j^*)$ where $g_i' \neq g_i^*$. Furthermore, let $d_i' > d_i^*$, $s_i' = s_i^* = s_j^* = s$. If $g_i^*$ and $g_j^*$ are strategies of a unique NE under EDCF using FLB, the equilibrium may be undesirable. And as a result, the unique SPE is also undesirable.

*Proof:* Using a similar procedure described in the proof of Lemma 7, we have

$$\frac{\gamma(g_j^*)}{\gamma(g_i')} * \frac{b_i' * s * \gamma(g_i') + b_j^* * s * \gamma(g_j^*) + t_{idle} * \gamma(g_i') * \gamma(g_j^*)}{b_i^* * s * \gamma(g_i^*) + b_j^* * s * \gamma(g_j^*) + t_{idle} * \gamma(g_i^*) * \gamma(g_j^*)} > \frac{b_i'}{b_i^*} * \frac{\alpha_i(g_i')}{\alpha_i(g_i^*)}$$

$$\frac{\gamma(g_j^*)}{\gamma(g_i')} * \frac{b_i^* * b_i' * s * \gamma(g_i') + b_i^* * b_j^* * s * \gamma(g_j^*) + b_i^* * t_{idle} * \gamma(g_i') * \gamma(g_j^*)}{b_i' * b_i^* * s * \gamma(g_i^*) + b_i' * b_j^* * s * \gamma(g_j^*) + b_i' * t_{idle} * \gamma(g_i^*) * \gamma(g_j^*)} > \frac{\alpha_i(g_i')}{\alpha_i(g_i^*)}$$

For $\frac{\alpha_i(g_i')}{\alpha_i(g_i^*)} > \frac{\gamma(g_j^*)}{\gamma(g_i')}$ (and hence $R^{prac}(g_i') > R^{prac}(g_i^*)$), it must be that the second term on the left is $> 1$. For instance, this will happen if $b_i^* \geq b_i'$ although it could still hold with $b_i^* < b_i'$ since $\gamma(g_i') > \gamma(g_i^*)$.

Intuitively, if node $i$ transmits at high data rate $d_i'$ and the loss rate experienced is high, node $i$ will not be able to transmit the maximum number of frames allowed under $t^{max}$. Therefore, if node $i$, by transmitting at a lower data rate $d_i^*$, can reduce the loss rate low enough such that $b_i^*$ is larger than $b_i'$, the node will prefer to use $d_i^*$ over $d_i'$ even though $R^{prac}(g_i') > R^{prac}(g_i^*)$.